\documentclass[11pt]{article}
\usepackage{amssymb}
\usepackage{amsmath}
\usepackage{amsfonts}

\newcommand{\be}{\begin{equation}}
\newcommand{\ee}{\end{equation}}
\newcommand{\bea}{\begin{eqnarray}}
\newcommand{\eea}{\end{eqnarray}}
\newcommand{\beano}{\begin{eqnarray*}}
\newcommand{\enano}{\end{eqnarray*}}

\newcommand{\Leb}[2]{
  \ifnum #2=1
  L^#1(\R,\,\ud x)
  \fi
  \ifnum #2=2
  L^#1(\R^2,\,\ud^2 \vec{x}\,)
  \fi}

\newcommand{\ud}{\mathrm{d}}

\begin{document}
\title{ Dissipative dynamics of the semiconductor-cavity QED with $q$-deformed bosons in the dispersive approximation}
\author{E. Karimzadeh Esfahani\thanks{E-mail
address: karimzadehehsan@gmail.com} , and R.
Roknizadeh\thanks{E-mail address: rokni@sci.ui.ac.ir}, and M. H.
Naderi\thanks{E-mail address: mhnaderi2001@yahoo.com}
~ \\ \\
{\it Quantum Optics Group, Department of Physics, University of Isfahan,}\\
{\it Isfahan, Iran}}
\date{ }
\maketitle
\begin{abstract}
In this paper we give fully analytical description of the
dynamics of a collection of $N$-Frenkel excitons in high density
regime dispersively coupled to a single mode cavity field, in the
presence of both exciton and cavity-field dissipations. By using
excitonic operators as $q$-deformed bosonic operators for the
system, we solve analytically the Liouville equation for the
density operator at zero temperature and investigate the
influence of the number of excitons and the effect of both
dissipations on dynamical behavior of the system. We use the
solution of master equation to explore the dissipative dynamics
of non-classical properties such as, molecule-field entanglement,
quadrature squeezing of the field, and molecular dipole squeezing.
We find that the non-classical properties are strongly affected by
the number of excitons and also by the existence of both
dissipations.
\end{abstract}
\section{Introduction}
In quantum theory and cavity-quantum electrodynamics (Cavity-QED) of
semiconductors where cavity is chosen as a bulk or confined systems
such as quantum wells and quantum dots, we work in the low density
regime of excitons, and treat these quasiparticles as ideal bosons
to solve wide number of problems in quantum optics, cavity-QED and
quantum information \cite{koch,yama}. Typically, these studies have
been done under the condition that the influence of the environment
is not taken into account. The environment which is represented by a
thermal reservoir always exist, and affects the system under
consideration. No matter how weak the coupling to such an
environment, the evolution of quantum subsystems is eventually
affected by non-unitary features such as decoherence, dissipation
and heating. The initial information irreversibly leaks out the
system into the very large number of uncontrollable degrees of
freedom of the environment. Initially prepared pure states are
typically corrupted on extremely short time scales due to quantum
coherence loss that turns them into mixed states \cite{22}. The
disspersive effects caused by the energy exchange between the system
and environment have been studied in the Jaynes-Cummings model (JCM)
\cite{Jaynes}, both analytically \cite{23} and numerically
\cite{24}. In the last few years the JCM with phase damping, as
applied to decoherence and entanglement \cite{25}, dissipative
dynamics of JCM model \cite{peixo}, and nonlinear quantum
dissipation effects on dynamical properties of the f-deformed JCM
model \cite{Naderi} have been studied. In addition to the exact
solvability of JCM within the rotating wave approximation, one of
the most interesting aspects of its dynamics is the entanglement
between atom and the field. Entanglement as a physical resource has
been used in quantum information science such as quantum
teleportation \cite{tele}, super dense coding \cite{sup} and quantum
cryptography \cite{cryp}. It has been shown that due to the
influence of the field dissipation in JCM under dispersive
approximation that the amplitude of the entanglement between the
field and the cavity-field decreases with the time and at least
completely suppresses \cite{peixo,Naderi,Zhu}. Sub-Poissonian
statistics and quadrature squeezing of the cavity-field are two
remarkable non-classical effects have also been studied in different
versions of generalized JCMs \cite{Naderi,dodonov}.

Fascinating electronic and optical properties of spatially
confined nanostructures like quantum wells and great potentially
of such structures in semiconductor cavity-QED application has
motivated permanent extension of their study. Among a variety of
new results in this field, it is important to find out the
influence of dissipation on non-classical properties, so in our
model, we replace one atom in JCM with $N$-Frenkel excitons.
However, if we want to replace two-level atoms in the cavity with
high density approximation of excitons in semiconductor
nanostructures, we can't use the standard Dicke model because the
excitons are not more ideal bosons. One way to deal with the
problem is to replace ideal bosons with $q$-deformed ones in the
Dicke model \cite{Liu}. We also consider the influence of
dissipation between excitons and their reservoir, visualized as a
large number of harmonic oscillators, and also the influence of
cavity-field damping due to its coupling to the environment, on
the dynamics of the system. For this purpose, we use the
Liouville equation for the density operator to study the
dynamical behavior of a dissipative system composed of
$N$-Frenkel excitons ($N$-two level molecules) interacting with a
single mode cavity-field and we show the influence of
dissipations and the number of molecules on quantum optical
properties of system such as, entanglement between the
cavity-field and molecules, quadrature field squeezing and also
molecular dipole squeezing.

This paper is organized as follows, In Sec.2 we first introduce the
Hamiltonian of the system without considering any dissipation in the
presence of excitonic operators as $q$-deformed bosons. In Sec.3, we
give an analytical solution for the master equation to derive the
total density operator in the presence of both dissipations. In
Sec.4 we employ the analytic results obtained in section.3 to
investigate the influence of number of molecules and the effect of
both reservoirs on the dynamical properties of molecules and
cavity-field. Finally, we summarize our conclusion in Sec.5.
\section{The Hamiltonian of the system}
We consider our system as $N$ identical two-level molecules of
splitting $\omega_{eg}$ interacting with a single-mode quantized
cavity-field of frequency $\omega_{0}$. We assume that all molecules
have equivalent mode position, so they interact with the
cavity-field by the same coupling constant $g$. In the case of
relatively high density of molecules in the excited state the
Hamiltonian of the system can be written in terms of the exciton
operators $b_{q}, b_{q}^\dagger$ (as $q$-deformed bosonic
operators)\cite{Liu}
\begin{equation}\label{H}
 \hat H = \hat b_q^ {\dagger}  \hat b_q \omega _{eg}  + \hat a^
{\dagger}  \hat a\omega _0  + \sqrt N g\left( {\hat a^ {\dagger}
\hat b _q+ \hat b_q^ {\dagger} \hat a} \right),\hspace{0.5cm}
\hbar=1,
\end{equation}
where $\hat{a}$ and $\hat{a}^{\dagger}$ are the annihilation and
creation operators of the quantized cavity-field respectively,
and the $q$-deformed bosonic operators $ \hat b_q $ and $ \hat b_q
^{\dagger}$ satisfying the $q$-deformed commutation relation
\begin{equation}
[ \hat b_q ,\hat b_q ^{\dagger}]_{q}= \hat b_q \hat b_q
^{\dagger}-q\hat b_q ^{\dagger}\hat b_q =1, \hspace{0.5cm}
q=1-\frac{2}{N},
\end{equation}
are defined as $\hat b_q = {{\hat b_g^ +  \hat b_e }
\mathord{\left/
 {\vphantom {{\hat b_g^ {\dagger}  \hat b_e } {\sqrt N }}} \right.
 \kern-\nulldelimiterspace} {\sqrt N }}$, $\hat b_q ^{\dagger} = {{\hat b_g \hat b_e^ +  } \mathord{\left/
 {\vphantom {{\hat b_g \hat b_e^ {\dagger}  } {\sqrt N }}} \right.
 \kern-\nulldelimiterspace} {\sqrt N }}$, in which $ \hat b_e , \hat b^{\dagger}_g $ $( \hat b_e^{\dagger}, \hat
b_g)$ are the annihilation and creation (creation and annihilation)
operators of one molecule in the excited state and the ground state,
respectively \cite{Liu}. Following Peixoto et all, \cite{peixo}, in
the large detuning approximation (dispersive limit), that is
\begin{equation}
\frac{\left|\delta\right|}{g}\gg\sqrt{n_{ph}+1}, \hspace{0.5cm}
\delta=\omega_{eg} - \omega_0,
\end{equation}
for any relevant photon number $n_{ph}$ , we obtain the following
effective interaction Hamiltonian
\begin{equation}\label{hef}
H_{eff}^{I}  = \omega A_{m,n} \left\{ {\left| {n,m} \right\rangle
\left\langle {n,m} \right|(\hat a^ {\dagger}  \hat a + 1) + \hat
a^ {\dagger}  \hat a\left| {n - 1,m + 1} \right\rangle
\left\langle {n - 1,m + 1} \right|} \right\},
\end{equation}
where we have defined $A_{m,n} \equiv n(m + 1) $, and $\omega
 \equiv {{g^2 } \mathord{\left/ {\vphantom {{g^2 } \delta}}\right.
\kern-\nulldelimiterspace} \delta }$. The state ${\left| {n,m}
\right\rangle }$ indicates that $n$ molecules ($m = N - n$
molecules) are in the excited state (ground state). Furthermore,
we have used
\begin{eqnarray}
 \hat b_q\left| {n,m} \right\rangle  = \frac{1}{{\sqrt N }}   \hat b_e  \hat b^{\dagger}_g \left| {n,m} \right\rangle =\frac{ \sqrt {n(m + 1)}}{{\sqrt N }}  \left| {n - 1,m + 1} \right\rangle , \\
 \hat b^ {\dagger}_q  \left| {n,m} \right\rangle  =  \frac{1}{{\sqrt N }}  \hat b_e^{\dagger} \hat b_g \left| {n,m} \right\rangle = \frac{\sqrt {(n + 1)m}}{\sqrt N }  \left| {n + 1,m - 1} \right\rangle .
 \\\nonumber
 \end{eqnarray}
The effective Hamiltonian (\ref{hef}) which does not cause any
transition in the system creates an entanglement between the
molecular and the field states.
\section{The master equation and its analytical solution}
We assume that our system is coupled with two types of reservoirs,
one of them is coupled to the cavity-field and the other is
coupled to molecules. The time evolution of the collection of $N$
two-level molecules interacting dispersively with a single mode
cavity-field in the interaction picture can be described by the
following master equation
\begin{equation}\label{mas}
\frac{d}{dt}\hat{\rho}(t)= -i\left[ H_{eff}^I , \hat{\rho}(t)
\right] + \hat{D}_{molecules} \hat{\rho}(t)+
\hat{D}_{field}\hat{\rho}(t),\hspace{0.5cm} \hbar=1,
\end{equation}
where $\hat{\rho}(t)$ is the density operator of the system, and
belongs to the set $D({\cal{H}}_M \otimes {\cal{H}}_F)$ of the
trace class operators acting in the space corresponding to the
direct product of the two Hilbert spaces ${\cal{H}}_M$ and
${\cal{H}}_F$ of the molecules and the field, $H_{eff}^I$ is
given by (\ref{hef}), and the damping of the cavity-field and
molecular subsystems are phenomenologically represented by the
superoperators $\hat{D}_{field}$ and $\hat{D}_{molecules}$. By
using thermal reservoir at zero temperature for the field and one
atom \cite{35,43} and the same one for molecules, using
$q$-deformed bosonic operators instead of atomic operators, we
can write these superoperators as
\begin{eqnarray}
\hat{D} _ {field} \cdot &=& k (2 \hat{a} \cdot \hat{a} ^ {\dagger}
- \hat{a} ^ {\dagger} \hat{a} \cdot - \cdot \hat{a}^{\dagger} \hat{a}),\\
\hat{D} _ {molecules} \cdot &=& k' (2 \hat b_q \cdot \hat
b^{\dagger}_q - \hat b^{\dagger}_q \hat b_q \cdot  - \cdot\hat
b^{\dagger}_q \hat b_q),
\end{eqnarray}
where $k$ and $k'$ are the corresponding dissipation constants.
These superoperators are linear combinations of bosonic
superoperators, and form a finite Lie algebra under commutation.
The bosonic superoperators represent the action of creation and
annihilation operators of the harmonic oscillator on an operator
$\hat{ O}$ :
\begin{equation}
 \begin{array}{l}
  (\hat a.)\hat O \equiv a^{\ell} \hat O \equiv \hat a\hat O,
  \hspace{0.5cm} (\hat a^ \dagger  .)\hat O \equiv (a^{\ell} )^
  \dagger \hat O
  \equiv \hat a^ \dagger \hat O,\\
  (.\hat a)\hat O \equiv a^r \hat O \equiv \hat O\hat a,
  \hspace{0.5cm} (.\hat a^ \dagger  )\hat O \equiv (a^r )^ \dagger
  \hat O \equiv \hat O\hat a^ \dagger.
 \end{array}
\end{equation}
Similarly, the action of $q$-deformed creation and annihilation
operators $ \hat b_q ^{\dagger}$ and $\hat b_q $ on an operator
$\hat O$ can be defined as
\begin{equation}
 \begin{array}{l}
  (\hat b_q.)\hat O \equiv b_q^{\ell} \hat O \equiv \hat b_q\hat O,
  \hspace{0.5cm} (\hat b_q^ \dagger  .)\hat O \equiv (b_q^{\ell} )^
  \dagger \hat O
  \equiv \hat b_q^ \dagger \hat O,\\
  (.\hat b_q)\hat O \equiv b_q^r \hat O \equiv \hat O\hat b_q,
  \hspace{0.5cm} (.\hat b_q^ \dagger  )\hat O \equiv (b_q^r )^ \dagger
  \hat O \equiv \hat O\hat b_q^ \dagger.
 \end{array}
\end{equation}
To solve the master equation (\ref{mas}), we rewrite it in the
basis of molecular states ${\left| {n,m} \right\rangle }$ (note
that there is just one mode of the cavity-field in the cavity
which can interact with one molecule)
\begin{equation}\label{mat}
\left( {\begin{array}{*{20}c}
   {\dot {\hat{\rho}} _{n,n} } & {\dot {\hat{\rho}} _{n,n - 1} }  \\
   {\dot  {\hat{\rho}} _{n - 1,n} } & {\dot  {\hat{\rho}} _{n - 1,n - 1} }  \\
\end{array}} \right) = \left( {\begin{array}{*{20}c}
   {\hat X_{n,n} } & {\hat X_{n,n - 1} }  \\
   {\hat X_{n - 1,n} } & {\hat X_{n - 1,n - 1} }  \\
\end{array}} \right).
\end{equation}
where $\dot \rho (t) \equiv \frac{d}{d t}\rho(t)  $, and we have
defined
\begin{subequations}
 \begin{eqnarray}
 &&\rho _{n,n} (t) \equiv \left\langle {n,m} \right|\hat \rho (t)\left| {n,m} \right\rangle , \\
 &&\rho _{n,n - 1} (t) \equiv \left\langle {n,m} \right|\hat \rho (t)\left| {n - 1,m + 1} \right\rangle , \\
 &&\rho _{n - 1,n} (t) \equiv \left\langle {n - 1,m + 1} \right|\hat \rho (t)\left| {n,m} \right\rangle , \\
 &&\rho _{n - 1,n - 1} (t) \equiv \left\langle {n - 1,m + 1} \right|\hat \rho (t)\left| {n - 1,m + 1} \right\rangle ,
 \\\nonumber
 \end{eqnarray}
\end{subequations}
Furthermore
\begin{subequations}\label{above}
 \begin{eqnarray}
  &&\hat X_{n,n}= \left\{ {i A_{m,n} \omega (P - M) - 2B_{n,m}^2 k' + k(2J - M - P)} \right\} \hat \rho _{n,n} (t)  ,\\
  &&\hat X_{n,n - 1}= \left\{ { - i A_{m,n} \omega (M + P + 1) - B_{n,m}^2 k' + k(2J - M - P)} \right\} \hat \rho _{n,n - 1}(t) ,  \\
  &&\hat X_{n - 1,n}= \left\{ {i A_{m,n} \omega (M + P + 1) - B_{n,m}^2 k' + k(2J - M - P)} \right\}\hat \rho _{n - 1,n} (t) ,
  \\\nonumber
  &&\hat X_{n - 1,n - 1}= \left\{ {i A_{m,n} \omega (M - P) + k(2J - M - P)} \right\} \hat \rho _{n - 1,n - 1} (t)  + 2B_{n,m}^2 k'\hat \rho _{n,n}(t)  ,
  \\
 \end{eqnarray}
\end{subequations}
are operators that act in ${\cal{H}}_F$ (remember that $n$ shows
the number of excited molecules). In Eqs. (\ref{above}), we have
defined $B_{n,m} \equiv {{\sqrt {n(m + 1)} } \mathord{\left/
 {\vphantom {{\sqrt {n(m + 1)} } {\sqrt N }}} \right.
 \kern-\nulldelimiterspace} {\sqrt N }} ={{\sqrt {A_{m,n}} } \mathord{\left/
 {\vphantom {{\sqrt {A(m,n)} } {\sqrt N }}} \right.
 \kern-\nulldelimiterspace} {\sqrt N }}$, and also $ M \equiv \hat a^ {\dagger} \hat a
\cdot $ , $ P \equiv \cdot \hat a^ {\dagger} \hat a$ , $J \equiv
\hat a \cdot \hat a^ {\dagger}$, which satisfy the following
commutation relations
\begin{equation}
\left[J, M \right]=J, \hspace{0.5cm} \left[J, P
\right]=J,\hspace{0.5cm} \left[M, P \right] = 0.
\end{equation}
In order to find the total density operator we need to solve the
four Liouvillian equations associated with each matrix element. For
any initial state, the total density operator can be evaluated from
the solution of Eq (\ref{mat}). We assume that the initial state of
the interacting system is given by
\begin{equation}\label{ini}
\left| {\psi _i } \right\rangle  = \frac{1}{{\sqrt 2 }}\left\{
{\left| {n,m} \right\rangle  + \left| {n - 1,m + 1} \right\rangle
} \right\} \otimes \left| \alpha  \right\rangle,
\end{equation}
which means that at $t=0$, molecules are in the superposition of
two states $\left| {n,m} \right\rangle$ and $\left| {n-1,m+1}
\right\rangle$, and the cavity-field is prepared in the Glauber
coherent state. After some calculation we evaluate the matrix
elements of the field density operator, using initial condition
(\ref{ini}), as
\begin{subequations}\label{dd}
 \begin{eqnarray}
  &&\hat \rho _{n,n} (t) = \frac{1}{{2}}e^{ - 2B_{n,m}^2 k't} \left| {\alpha e^{ - kt} e^{ - i  A_{m,n} \omega t} } \right\rangle \left\langle {\alpha e^{ - kt} e^{ - i A_{m,n} \omega t} } \right| \\
  &&\hat \rho _{n,n - 1} (t) = \frac{1}{{2}}e^{ - B_{n,m}^2 k't} e^{i\theta (t) + \Gamma (t)} \left| {\alpha e^{ - kt} e^{ - i A_{m,n} \omega t} } \right\rangle \left\langle {\alpha e^{ - kt} e^{i A_{m,n} \omega t} } \right| \\
  &&\hat \rho _{n - 1,n} (t) = \frac{1}{{2}}e^{ - B_{n,m}^2 k't} e^{ - i\theta (t) + \Gamma (t)} \left| {\alpha e^{ - kt} e^{i A_{m,n} \omega t} } \right\rangle \left\langle {\alpha e^{ - kt} e^{ - i A_{m,n} \omega t} } \right|
  \\\nonumber
  &&\hat \rho _{n - 1,n - 1} (t) = \frac{1}{{2}}\left| {\alpha e^{ - kt} e^{i A_{m,n} \omega t} } \right\rangle \left\langle {\alpha e^{ - kt} e^{i A_{m,n} \omega t} } \right| + \frac{1}{{2}}e^{ - \left| \alpha  \right|^2 e^{ - 2kt} }
  \\\nonumber
  {\rm    } &&\times \sum\limits_{n,n'} {\frac{{(\alpha e^{ - kt} e^{i A_{m,n} \omega t} )^n }}{{\sqrt {n!} }}} \frac{{(\alpha ^ *  e^{ - kt} e^{ - iA_{m,n} \omega t} )^{n'} }}{{\sqrt {n'!} }}\left| n \right\rangle \left\langle {n'} \right|\frac{{2B_{n,m}^2 k'}}{{ - 2B_{n,m}^2 k' + 2i A_{m,n} \omega (n' -
  n)}}\\\nonumber
  &&\times \left( {e^{ - 2B_{n,m}^2 k't} e^{2i A_{m,n} \omega t(n' - n)}  - 1} \right)  \\
 \end{eqnarray}
\end{subequations}
where the definitions
 \begin{eqnarray}\nonumber
 &\Gamma (t)  &=  - \left| \alpha  \right|^2 (1 - e^{ - 2kt} ) \\
 &&- \frac{{\left| \alpha  \right|^2 k}}{{k^2  + A_{m,n} ^2 \omega ^2 }}\left(
 {e^{ - 2kt} (k\cos (2 A_{m,n} \omega t) -  A_{m,n} \omega \sin( 2 A_{m,n} \omega t)) - k}
 \right),\\\nonumber
 &\Theta (t)  &=  -  A_{m,n} \omega t + \frac{{\left| \alpha
 \right|^2
 k}}{{k^2  +  A_{m,n} ^2 \omega ^2 }}\\
 &&  \times\left( {e^{ - 2 k t} (k \sin (2 A_{m,n} \omega t )+
  A_{m,n} \omega \cos (2 A_{m,n} \omega t)) -  A_{m,n} \omega } \right)\\\nonumber
 \end{eqnarray}
have been used. As a result, the total density operator of the
interacting system can be written as
\begin{equation}\label{den}
 \begin{array}{l}
  \hat \rho _{total}(t) = \hat \rho _{n,n} (t) \otimes \left| {n,m}
  \right\rangle \left\langle {n,m} \right| + \hat \rho _{n,n - 1}
  (t) \otimes \left| {n,m} \right\rangle \left\langle {n - 1,m + 1}
  \right|\\
  + \hat \rho _{n - 1,n} (t) \otimes \left| {n - 1,m + 1}
  \right\rangle \left\langle {n,m} \right| + \hat \rho _{n - 1,n -
  1} (t) \otimes \left| {n - 1,m + 1} \right\rangle \left\langle {n
  - 1,m + 1} \right|.
 \end{array}
\end{equation}
By taking the trace of the total density operator (\ref{den}) on
the molecular (field) variables, we get the reduced field
(molecular) density operator: $\hat \rho_f (t)= Tr_{molecule}
\left(\hat \rho_{total} (t)\right)$ ( $\hat \rho_m (t)=
Tr_{field} \left(\hat \rho_{total} (t)\right)$). In the next two
sections we are going to use the solution given by (\ref{den}) to
investigate the dynamical properties of the system.
\section{Dynamical properties of the model}
\subsection{Linear entropies and molecules-field entanglement}
It is well-known that if the cavity-field and molecules are
initially prepared in a pure state, then at $t>0$ the
molecules-field system evolves into an entangled state. In this
entangled state the field and molecules are separately in mixed
states. The stability of a pure state may be understood as the
process where quantum coherence of the state is preserved along
its time evolution. In this sense we say that an initial pure
quantum state, described by the density operator $\hat \rho(t)$
is stable if $Tr \hat \rho^2(t)=1$, for all times. One way to
measure the stability of an initial pure state is to use the
linear entropy \cite{38}
\begin{equation}\label{entrop}
s = 1 - Tr \hat \rho^2(t).
\end{equation}
The time evolution of the molecule(field) entropy reflects the
time evolution of the degree of entanglement between the
molecules and the field. The higher the entropy is, the greater
the entanglement between the molecule and the field becomes. By
using Eqs.(\ref{den}) and (\ref{entrop}) the linear entropy of the
total system under consideration is obtained as follows
\begin{equation}\label{et}
 \begin{array}{l}
  s_{m - f} (t) = 1 - Tr\left( {\hat \rho _{total}^2(t) } \right) = \frac{1}{4}(1 + e^{ - 4B_{n,m}^2 k't}  + 2e^{2\Gamma } e^{ - 2B_{n,m}^2 k't}  \\
  + 2e^{ - 2\left| \alpha  \right|^2 e^{ - 2kt} } \sum\limits_{n,n'} {\frac{{\left( {\left| \alpha  \right|^2 e^{ - 2kt} } \right)^{n + n'} }}{{n!n'!}}} \frac{1}{{( - 2B_{n,m}^2 k')^2  + 4 A_{m,n}^2 \omega ^2 (n - n')^2 }} \\
  \times ( ( - (2B_{n,m}^2 k')^2 (e^{ - 2B_{n,m}^2 k't} \cos(2 A_{m,n} \omega (n - n')t) - 1) \\
  + 4 A_{m,n} \omega B_{n,m}^2 k'(n - n')e^{ - 2B_{n,m}^2 k't} \sin(2 A_{m,n} \omega (n - n')t)) ) \\
  + e^{ - 2\left| \alpha  \right|^2 e^{ - 2kt} } \sum\limits_{n,n'} {\frac{{\left( {\left| \alpha  \right|^2 e^{ - 2kt} } \right)^{n + n'} }}{{n!n'!}}} \frac{{(2B_{n,m}^2 k')^2 }}{{( - 2B_{n,m}^2 k')^2  + 4 A_{m,n}^2 \omega ^2 (n - n')^2 }} \\
  \times \left\{ {e^{ - 4B_{n,m}^2 k't}  - 2e^{ - 2B_{n,m}^2 k't} \cos(2 A_{m,n} \omega (n - n')t) + 1} \right\}). \\
 \end{array}
\end{equation}
Furthermore, the linear entropy of the cavity-field is given by
\begin{equation}\label{ef}
\begin{array}{l}
 s_f (t) = 1 - Tr_f \left( {\hat \rho _f^2 (t)} \right) = 1 + e^{ - 4B_{n,m}^2 k't}  + 2e^{ - 2B_{n,m}^2 k't} e^{2\left| \alpha  \right|^2 e^{ - 2kt} (\cos (2 A_{m,n} \omega t) - 1)}  \\
  + e^{ - 2\left| \alpha  \right|^2 e^{ - 2kt} } \sum\limits_{n,n'} {\frac{{(\left| \alpha  \right|^2 e^{ - 2kt} )^{n + n'} }}{{n!n'!}}\frac{{(2B_{n,m}^2 k')^2 }}{{(2B_{n,m}^2 k')^2  + 4 A_{m,n}^2 \omega ^2 (n - n')^2 }}}  \\
  \times \left\{ {e^{ - 4B_{n,m}^2 k't}  - 2e^{ - 2B_{n,m}^2 k't} \cos (2 A_{m,n} \omega (n' - n)t) + 1} \right\} \\
  + 2(e^{ - 2B_{n,m}^2 k't}  + 1)e^{ - 2\left| \alpha  \right|^2 e^{ - 2kt} } \sum\limits_{n,n'} {\frac{{(\left| \alpha  \right|^2 e^{ - 2kt} )^{n + n'} }}{{n!n'!}}} \frac{1}{{(2B_{n,m}^2  k')^2  + 4 A_{m,n}^2 \omega ^2 (n - n')^2 }} \\
  \times (( - (2B_{n,m}^2 k')^2 \left( {e^{ - 2B_{n,m}^2 k't} \cos (2 A_{m,n} \omega (n' - n)t) - 1} \right) \\
  + 4 A_{m,n}^2 \omega ^2 B_{n,m}^2 k'(n' - n)e^{ - 2B_{n,m}^2 k't} \sin (2 A_{m,n} \omega (n' - n)t) )), \\
 \end{array}
\end{equation}
and the molecular coherence loss will be measured by
\begin{equation}\label{em}
\begin{array}{l}
s_m (t) = 1 - Tr_m \left( {\hat \rho _m^2 (t)} \right) \\
= \frac{1}{4}(e^{ - 4B_{n,m}^2 k't}  + 2e^{2\Gamma } e^{ -
2B_{n,m}^2 k't} e^{\left| \alpha  \right|^2 e^{ - 2kt} (2\cos
(2A_{n,m} \omega t) - 2)}
+ 4 + e^{ - 4B_{n,m}^2 k't}  - 4e^{ - 2B_{n,m}^2 k't} ). \\
\end{array}
\end{equation}
According to Eqs.(\ref{et})-(\ref{em}), we plot the time
evolution of the linear entropy of the total system, the
cavity-field and the molecular subsystem as functions of scaled
time $\omega t$, in Figs. 1.

First, we discuss the coherence loss of the field. For all diagrams
in Figs. 1, the field is initially prepared in the coherence state
and $s_f(t)=0$, but in a short stage $s_f(t)$ increases and the
coherence of the field loses (of course the equilibrium state of the
filed correspond to the vacuum, for which the linear entropy of the
field is zero). The linear entropy of the field $s_f$ shows local
maxima and minima, corresponding to the entanglement and
disentanglement between the field and excitons. In the presence of
molecular dissipation (Figs. 1a, 1b), the field linear entropy still
exhibits local maxima and minima, but minimum of disentanglement is
not taking place at $s_f=0$ and the field is not in a pure state. In
the absence of molecular dissipation, in figure.1c, when
disentanglement happens periodically, the field is in a pure state
($s_F=0$).

To verify the role of two types of dissipations on $s_m$ and
$s_{m -f}$, we discuss the coherence loss of the molecules and of
the system. We find that the linear entropy of the system and of
molecules exhibit periodic behavior, and these periods coincide
with the related periods of $s_f$. Note that the field and
molecules are not in pure state. In the course of time evolution,
time, both $s_m (t)$ and $s_{m-f}(t)$, in the presence of
molecular dissipation (Figs. 1a, 1b), find the maximum values
(corresponding to maximum coherence loss ), and then these linear
entropies decrease smoothly to find their asymptotic values at
zero (the equilibrium state of molecules corresponds to the state
that all molecules are in the ground state and the linear
enrtropies are zero). Furthermore, in the absence of molecular
dissipation, these two entropies tend to an asymptotic non zero
value (Figs. 1c, 1d). We also show the linear entropies for one
molecule in the absence of molecular dissipation (Fig. 1d), and
find the same results as the usual JCM given in reference
\cite{peixo}.
\subsection{Quadrature squeezing of the cavity-field}
In the past two decades, there has been major studies focused on
the fluctuations in the quadrature amplitude of the
electromagnetic field to produce squeezed light. This light is
indicated by having less noise in one field quadrature than
vacuum state with an excess of noise in the conjugate quadrature
such that the product of canonically conjugate variances must
satisfy the uncertainty relation. In the studies of quantum
optics theory, this light occupies a wide area because of the
various applications, e.g., in optical communication networks
\cite{40}, in interferometric techniques \cite{41}, and in
optical waveguide tap \cite{42}. Furthermore, investigation of
the squeezing properties of the radiation field is a central
topics in quantum optics and noise squeezing can be measured by
means of homodyne detection \cite{43}.

In order to investigate the quadrature squeezing of the model
under consideration, we introduce the two slowly varying
Hermition operators $\hat X_{1a} (t)$ and $\hat X_{2a} (t)$
defined by, respectively,
\begin{equation}
\hat X_{1a} (t) = \frac{1}{2}\left( {\hat a e^{i\omega _0 t}  +
\hat a^ {\dagger} e^{ - i\omega _0 t} } \right), \hspace{0.5cm}
\hat X_{2a} (t) = \frac{1}{2}\left( {\hat ae^{i\omega _0 t} -\hat
a^ {\dagger} e^{ - i\omega _0 t} } \right).
\end{equation}
A state of the field said to be squeezed when one of the
quadrature components $\hat X_{1a} (t)$ and $\hat X_{2a} (t)$
satisfies the relation
\begin{equation}
\left\langle {\left( {\Delta \hat X_{ia} (t)} \right)^2 }
\right\rangle  < \frac{1}{4}, \hspace{0.5cm} (i=1 or 2).
\end{equation}
The degree of squeezing can be measured by the squeezing
parameter $s_i (i=1,2)$ defined by
\begin{equation}
s_i (t) \equiv 4\left\langle {\left( {\Delta \hat X_{ia} (t)}
\right)^2 } \right\rangle  - 1 .
\end{equation}
Therefore the condition for squeezing in the quadrature component
can be simply written as $s_i (t)< 0 $. In Figs. 2, we have
plotted the squeezing parameter $ s_1 (t)$ given by
\begin{equation}
\begin{array}{l}
 s_1 (t) = 2e^{ - 2\left| \alpha  \right|^2 e^{ - 2kt} } \sum \limits_n {\frac{{(\left| \alpha  \right|^2 e^{ - 2kt} )^n }}{{n!}}} n + 2\{ e^{ - 2B_{n,m}^2 k't} \left| \alpha  \right|^2 e^{ - 2kt} \cos (2(\omega _0  -  A_{m,n} \omega )t) \\
  + \left| \alpha  \right|^2 e^{ - 2kt} \cos (2(\omega _0  + A_{m,n} \omega )t)\frac{1}{{(2B_{n,m}^2 k')^2  + 4A_{m,n} ^2 \omega ^2 ( - 2)^2 }} \\
  \times \left\{ { - (2B_{n,m}^2 k')^2 \left( {e^{ - 2B_{n,m}^2 k't} \cos (4 A_{m,n} \omega t) - 1} \right) + 8 A_{m,n} ^2\omega ^2 B_{n,m}^2 k'e^{ - 2B_{n,m}^2 k't} \sin (4 A_{m,n} \omega t)} \right\} \\
  + \left| \alpha  \right|^2 e^{ - 2kt} \sin (2(\omega _0  + A_{m,n} \omega )t)\frac{1}{{(2B_ {n,m}^2 k')^2  + 4A_{m,n} ^2\omega ^2 ( - 2)^2 }} \\
  \times \left\{ { - 8 A_{m,n} ^2 \omega ^2 B_{n,m}^2 k'\left( {e^{ - 2B_{n,m}^2 k't} \cos (4 A_{m,n} \omega t) - 1} \right) - (2B_{n,m}^2 k')^2 e^{ - 2B_{n,m}^2 k't} \sin (4 A_{m,n} \omega t)} \right\}\}  \\
  + 4\{ e^{ - 2B_{n,m}^2 k't} \alpha e^{ - kt} \cos ((\omega _0  -  A_{m,n} \omega )t) \\
  + \alpha e^{ - kt} \cos ((\omega _0  + A_{m,n} \omega )t)\frac{1}{{(2B_{n,m}^2 k')^2  + 4 A_{m,n} ^2\omega ^2 }} \\
  \times \left\{ { - (2B_{n,m}^2 k')^2 \left( {e^{ - 2B_{n,m}^2 k't} \cos (2 A_{m,n} \omega t) - 1} \right) + 4 A_{m,n} ^2 \omega ^2 B_{n,m}^2 k'e^{ - 2B_{n,m}^2 k't} \sin (2 A_{m,n} \omega t)} \right\} \\
  + \alpha e^{ - kt} \sin ((\omega _0  + A_{m,n}  \omega )t)\frac{1}{{(2B_{n,m}^2 k')^2  + 4 A_{m,n} ^2 \omega ^2 }} \\
  \times \left\{ { - 4A_{m,n} ^2 \omega ^2 B_{n,m}^2 k'\left( {e^{ - 2B_{n,m}^2 k't} \cos (2 A_{m,n} \omega t) - 1} \right) - (2B_{n,m}^2 k')^2 e^{ - 2B_{n,m}^2 k't} \sin (2 A_{m,n} \omega t)} \right\}\}
  ^2,
  \\
 \end{array}
\end{equation}
versus the scaled time $\omega t$ for the corresponding data used
in Figs. 1. As it is seen, the squeezing parameter $s_1(t)$ shows
fast oscillations and the quadrature component $\hat X_{1a}$ does
not exhibit squeezing. In the presence of both dissipations (Fig.
2a), squeezing parameter $s_1(t)$ present local maxima and minima
due to the field interaction with molecules. In the presence of
molecular dissipation (Figs. 2a, 2b), the field is no longer in a
pure state, except initial time and equilibrium state
(corresponding to the vacuum), and in the absence of molecular
dissipation (Fig. 2c), periodically, the field can be find in pure
state.
\subsection{Molecular dipole squeezing} In this section we are going to discuss about the
dynamical behavior of excitonic ensemble property that is
molecular dipole squeezing.

To analyze the quantum fluctuation dipole variables, we consider
$\hat{\sigma}_x$ and $\hat{\sigma}_y$ corresponding to the
dispersive and absorptive components of the amplitude of molecular
polarization\cite{5.5}
\begin{equation}
\hat \sigma_x (t) = \frac{1}{2}(\hat b ^\dagger_q  e^{ - i\omega
_{eg} t} + \hat b _q  e^{i\omega _{eg} t} ), \hspace{0.5cm} \hat
\sigma_y (t) = \frac{1}{{2i}}(\hat b^\dagger_q  e^{ - i\omega
_{eg} t} - \hat b_q e^{i\omega _{eg} t} ).
\end{equation}
The fluctuations in the component $\hat{\sigma}_i$($i=x$ or $y$)
are said to be squeezed if the variance in $\hat{\sigma}_i$
satisfies the condition $\left\langle {\left( {\Delta \hat
\sigma_x (t)} \right)^2 } \right\rangle  < \frac{1}{4}\left|
{\left\langle {\hat \sigma_z (t)} \right\rangle } \right|$
$\hspace{0.15cm}$ ($i=x$ or $y$). Since $\left\langle {\hat
\sigma_i^2 (t)} \right\rangle  = 1/4$ the condition of dipole
squeezing may be written as
\begin{equation}
F_i (t) = 1 - 4\left\langle {\hat \sigma_i (t)} \right\rangle ^2
- \left| {\left\langle {\hat \sigma_z (t)} \right\rangle }
\right| < 0 \hspace{0.5cm}(i=x \hspace{0.1cm} or \hspace{0.1cm} y)
\end{equation}
We find $F_y (t)$, corresponding to the squeezing of
$\hat{\sigma}_y (t)$ for the system under consideration, as
follows
\begin{equation}\label{dipole}
F_y (t) = 1 - 4\left\{ {\frac{1}{2} B_{n,m} e^\Gamma  e^{ -
B_{n,m}^2 k't} e^{ - \left| \alpha  \right|^2 e^{ - 2kt} (\cos (2
A_{m,n} \omega t) - 1)} \sin ( \omega _{eg} t + \theta  - \left|
\alpha \right|^2 e^{ - 2kt} \sin (2 A_{m,n} \omega t))} \right\}^2.
\end{equation}
To investigate the influence of both kinds of dissipations, we
plot the time evolution of $F_y (t)$ in Figs. 3-6. As it is seen,
in the course of time evolution, the function $F_y(t)$ shows rapid
oscillations which are damped in the presence of dissipations,
and under the condition that $B_{n,m}\neq1$, dipole squeezing can
occur. In Figs. 3, we show the influence of dissipations on
$F_y(t)$, in the presence of both dissipations (figure 3a), in
the absence of field dissipation (figure 3b), and in the absence
of molecular dissipation (figure 3c). We can see that with
suppression of dissipations in Figs. 3b,3c, the rate of damping in
oscillations decreases, and in the absence of molecular
dissipation the oscillations continue to narrow band between two
positive numbers. In Figs. 4a-4c, we show the influence of total
number of excitons $N$ on $F_y (t)$. By increasing the number of
excitons the squeezing can be much stronger and the oscillations
damped faster.

Note that in all diagrams, under the condition $B_{n,m} \neq 1$,
dipole squeezing appears and for any total number of excitons $N$,
as it is clear in relation (\ref{dipole}), for larger numbers of
$B_{n,m}$, the effect of squeezing will be stronger. In Figs. 5a-5c,
we show the influence of the parameter $B_{n,m}$ on $F_y (t)$. As we
can see, with $B_{n,m} =1$ there is no squeezing for $F_y (t)$.
There is a an interesting instance for when molecular dissipation
can be suppressed that by increasing the number of molecules,
$F_{y}(t)$ shows squeezing all the time, periodically
(figure.6a,6b), and by increasing the number of molecules it will be
much stronger.
\section{Summary and Conclusions}
In this paper, we have studied theoretically the influence of the
dissipations on non-classical properties of the system as
$N$-Frenkel excitons interacting with a single mode cavity-field
in the dispersive approximation. Using $q$-deformed bosonic
operators for excitons, we obtained an effective interaction
Hamiltonian and we used it to find the master equation containing
molecular dissipation and cavity-field dissipation. By solving the
master equation, the total density operator was obtained and some
of non-classical properties were studied. In continue we summarize
our conclusions:

First, we discussed the linear entropy to find the coherence loss
of the cavity-field $s_f(t)$, molecules $s_m (t)$ and of the
system $s_{m-f}(t)$. We found that the linear entropies present
local maxima and minima corresponding to entanglement and
disentanglement due to the field interaction with excitons. As
the time goes on, in the presence of both dissipations, all
linear entropies decrease smoothly to find their asymptotic zero
values.

second, we discussed about quadrature squeezing of the field. We
found that in the presence of dissipations, squeezing parameter
$s_1(t)$ shows fast oscillations with local maxima and minima due
to the field interaction with molecules and the quadrature
component $\hat X_{1a}$ does not exhibit squeezing.

In the last section we discussed the influences of number of
molecules and both dissipations on the temporal evolution of
molecule dipole squeezing $F_y(t)$. We found that $F_y (t)$ shows
rapid oscillations between positive and negative values that the
minus values corresponding to the squeezing of $\hat{\sigma}_y
(t)$ and Squeezing can be occur when $B_{n,m}\neq1$ (the parameter
$B_{n,m}$ has introduced in section 3, is related to the number
of molecules which are in the excited and ground state), and so
there is no squeezing when we have just one molecule in the
cavity. We also found that by increasing the number of molecules,
the squeezing can be much stronger and oscillations of $F_y(t)$
damped faster.\\

\textbf{Acknowledgment} The authors wish to thank
      the Office of Graduate Studies of the University of Isfahan for
      their support. This work is also supported by Nanotechnology
      Initiative of Iran.

\newpage

\textbf{Figure 1.} Time evolution of the linear entropy
$s_{a-f}(t)$ (black solid curve), $s_f(t)$ (gray solid curve) and
$s_a (t)$ (dashed curve) as functions of scaled time $\omega t$,
for $\alpha = 1$, $N = 10$ and (Fig.1a) $k = k^{\prime} =
0.05\omega$, $n=m=5$ (Fig.1b); $k=0$, $k^{\prime} = 0.05\omega $
(Fig.1c); $k =0.05 \omega$ $k^{\prime} = 0$, $A_{n,m}=B_{n,m}=1$ (Fig.1d).\\
\\
\\

\textbf{Figure 2.} The time evolution of $s_1$ as a function of
scaled time $\omega t$, for $N=10$, $n=m=5$ $\alpha=1$, and for
$k =k^{\prime}= 0.05\omega$ (Fig.2a); $k= 0$,
$k^{\prime}=0.05\omega$ (Fig.2b); $k = 0.05\omega$, $k^{\prime}=0$
(Fig.2c).\\
\\
\\

\textbf{Figure 3.} The time evolution of $F_y(t)$ as a function of
scaled time $\omega t$, for $N=10$, $m,n=5$, $\alpha=1$, and when
$B_{n,m}\neq1$: in the presence of both dissipations $ k
=k^{\prime} = 0.05\omega $ (Fig.3a); in the absence of field
dissipation $k =0$, $k^{\prime}=0.05\omega$ (Fig.3b); in the
absence of
molecular dissipation $k=0.05\omega$, $k^{\prime}=0$ (Fig.3c).\\
\\
\\

\textbf{Figure 4.} The time evolution of $F_y(t)$ as a function of
scaled time $\omega t$ for three values of $N$, $ k =k^{\prime} =
0.05\omega $, $\alpha=1$ when $B_{n,m}\neq1$: $N=20$, $m=10$
(Fig.4a);
$N=50$, $m=25$ (Fig.4b); $N=100$, $m=50$ (Fig.4c).\\
\\
\\

\textbf{Figure 5.} The time evolution of $F_y(t)$ as a function of
scaled time $\omega t$ for $\alpha=1$, when $B_{n,m} = 1$: $N=1$
(Fig.5a); $N=10$ (Fig.5b); $N=100$ (Fig.5c).\\
\\
\\

\textbf{Figure 6.} The time evolution of $F_y(t)$ as a function
of scaled time $\omega t$, for $\alpha=1$, $k^{\prime} = 0$, $k
=0.05\omega$, when $B_{n,m}\neq1$: $N=30$, $m=15$ (Fig.6a);
$N=100$,
$m=50$ (Fig.6b).\\
\\
\\

\begin{thebibliography}{99}
 \bibitem{koch}H. Haug, S. W. Koch, Quantum Theory of The Optical and Electronic Properties
     of Semiconductors (Word Scientific Publishing Co. Ltd, 1993).
 \bibitem{yama}Y.Yamamoto, F. Tassone , and H. Cao, Semiconductor Cavity Quantum
     Electrodynamics (Schpringer-Verlag Berlin Heidelberg New York, 2000).
 \bibitem{22}R. Omnes, Phys. Rev. A {\bf 56}, 3383 (1997); W.
     H. Zurek, Phys. Rev. D {\bf 24}, 1516 (1981); H. Zurek, Phys.
     Today {\bf 44}, 36 (1991).
 \bibitem{Jaynes}E. T. Jaynes, F.W.Cummings, IEEE {\bf 51}, 89 (1963).
 \bibitem{23}S. M. Barnett, P. L. Knight, Phys. Rev. A {\bf 33},
     2444 (1986); R. R. Puri, G. S. Agarwal, Phys. Rev. A {\bf 35},
     3433 (1987); H. J. Briegel, B. G. Englert, Phys. Rev. A {\bf 47},
     3311 (1993);A. J. Wonderen, Phys. Rev. A {\bf 56}, 3116 (1997);
     A. Lindner, H. Freese, G. Quehl, D. Reiß, K. Schiller, Eur. Phys.
     J. D {\bf 17}, 99 (2001).
 \bibitem{24}J. Eiselt, H. Risken, Opt. Commun. {\bf 72}, 351
     (1989); J. Gea-Banachloce, Phys. Rev. A {\bf 47}, 2221 (1993); B.
     G. Englert, N. Naraschevski, D. Schezlr, Phys. Rev. A {\bf 50},
     2667 (1994).
 \bibitem{25}L. M. Kuang, X. Cheng, G. H. Chen, M. L. Ge, Phys. Rev. A {\bf 56},
     3139 (1997); H. A. Hessian, H. J. Ritsch, J. Phys. B: At. Mol.
     Opt. Phys. {\bf 35}, 4619 (2002).
 \bibitem{peixo}J. G. Peixoto de Faria, and M. C. Nemes, Phys. Rev. A{\bf
     59}, 3918 (1999).
 \bibitem{Naderi}M. H. Naderi and M. Soltanolkotabi, Eur. Phys. J. D {\bf
     39}, 471 (2006).
 \bibitem{tele}D. S. Freitas, A. Vidiella-Barranco, J. A. Roversi,
     Phys. Lett. A{\bf 249}, 275 (1998).
 \bibitem{sup}C. H. Bennett, G. Brassard, C. Cr\'{e}peau, R.
     Jozsa, A. Peres, and W. K. Wootters, Phys. Rev. Lett. {\bf 70},
     1895 (1993).
 \bibitem{cryp}C. H. Bennet, S. J. Wiesner, Phys. Rev. Lett. {\bf
     69}, 2881 (1992).
 \bibitem{Zhu}L. Zhou, H. S. Song, and Y. X. Luo, J. Opt. B:
     Quantum Semiclass. Opt. {\bf 4}, 103 (2002).
 \bibitem{dodonov}V. V. Dodonov, W. D. Jos\'{e}, and S. S. Mizrahi, J. Opt. B:
     Quantum Semiclass. Opt. {\bf 5}, 567 (2003); Le-Man Kuang, Xin Chen, Guang-Hong Chen, and Mo-Lin Ge
 , Phys. Rev. A{\bf 56}, 3139 (1997); Le-Man Kuang, Xin Chen, and Mo-Lin Ge
 , Phys. Rev. A {\bf52}, 1857 (1995).
 \bibitem{Liu}Yu-Xi Liu, C. P. Sun, S. X. Yu, and D. L. Zhou, Phys. Rev. A {\bf 63},
     023802 (2001).
 \bibitem{35}A. Royer, Phys. Rev. A{\bf
     43}, 44 (1991); A. Royer, Phys. Rev. A{\bf 45}, 793 (1992); S. J.
     Wang, M. C. Nemes, A. N. Salguero, H. A. Weidenmuller, Phys. Rev.
     A{\bf 66}, 033608 (2002).
 \bibitem{43}M. O. Scully, M. S. Zubairy, Quantum Optics (Cambridge University Press, Cambridge,
     1997).
 \bibitem{38}W. H. Zurek, S. Habib, J. P. Paz, Phys. Rev. Lett. {\bf
     70}, 1187 (1993); J. I. Kim, M. C. Nemes, A. F. R. de Toledo
     Piza, H. E. Borges, Phys. Rev. Lett. {\bf 77}, 207 (1996); A.
     Isar, A. Sandulescu, W. Scheid, Phys. Rev. E{\bf 60}, 6371 (1999).
 \bibitem{40}H. P. Yuen, J. H. Shapiro, IEEE Trans. Inform. Theory {\bf
     IT-24}, 657 (1978); J. H. Shapiro, H. P. Yuen, M. J. A. Machado,
     IEEE Trans. Inform. Theory {\bf IT-25}, 179 (1979); H. P. Yuen,
     J. H. Shapiro, IEEE Trans. Inform. Theory {\bf IT-26}, 78 (1980).
 \bibitem{41}C. M. Caves, B. L. Schumaker, Phys. Rev. A {\bf 31},
     3068 (1985); B. L. Schumaker, C. M. Caves, Phys. Rev. A {\bf 31},
     3093 (1985).
 \bibitem{42}J. H. Shapiro, Opt. Lett. {\bf 4}, 351 (1980)
 \bibitem{5.5}S. M. Barnett, Opt. Commun. {\bf 61}, 432 (1982); P.
     Zhou, J. S. Peng, Phys. Rev. A {\bf 44}, 3331 (1991).
\end{thebibliography}
\end{document}